\documentclass[3p,times]{elsarticle}

\usepackage{ecrc}

\volume{00}

\firstpage{1}

\journalname{Physics Letters B}

\runauth{}

\jid{procs}

\jnltitlelogo{Physics Letters B}

\CopyrightLine{2011}{Published by Elsevier Ltd.}

\usepackage{amssymb}

\usepackage[figuresright]{rotating}

\begin{document}

\begin{frontmatter}



\dochead{}

\title{On the sensitivity of extracting the astrophysical cross section factor of the $^{12}$C$(\alpha,\gamma)$ reaction from existing data.}


\author{Moshe Gai}

\address{LNS at Avery Point, University of Connecticut, Groton, Connecticut 06340-6097 \\
 and Wright Lab, Department of Physics, Yale University, New Haven, Connecticut 06520-8124}

\begin{abstract}
We address a conflicting report on the value and uncertainty of the astrophysical cross section factor of the $^{12}$C$(\alpha,\gamma)$ reaction extracted from existing data. In sharp contrast to previously reported ambiguities (by up to a factor 8), Schuermann {\em et al.} suggest an accuracy of 12\%. We demonstrate that the so claimed ``rigorous data selection criteria" used by Schuermann {\em et al.} relies on the s-factors extracted by Assuncao {\em et al.} But these results were shown in a later analysis (by this author) to have large error bars (considerably larger than claimed by Assuncao {\em et al.}) which render these data not appropriate for a rigorous analysis. When their ``rigorous data selection" is adjusted to remove the results of Assuncao {\em et al.} the astrophysical cross section factor cannot be extracted with 12\% accuracy, or even close to it. Such data on the $S_{E2}$ values at low energies deviate by up to a factor two from their fit and exhibit a sharper slope rising toward low energies, leading to strong doubt on their extrapolated $S_{E2}(300)$ value and the quoted small error bar. Contrary to their claim the small value of $S_{E1}(300) \approx 10$ keVb cannot be ruled out by current data including the most modern gamma-ray data. As previously observed by several authors current data reveal ambiguities in the value of $S_{E1}(300)$ = approximately 10 keVb or approximately 80 keVb, and the new ambiguity that was recently revealed (by this author) of $S_{E2}(300)$ = approximately 60 keVb or approximately 154 keVb, appear to be a more reasonable evaluation the status of current data.

\end{abstract}


\end{frontmatter}



\vspace{1cm}

During stellar helium burning oxygen is formed by the fusion of helium with carbon and is denoted by the $^{12}$C$(\alpha,\gamma)$ reaction. It has been announced three decades ago \cite{Fow84} as the most important and uncertain nuclear input of stellar evolution and it remains so today. In sharp contrast Schuermann {\em et al.} \cite{Sch12} claimed that the astrophysical cross section factor of the $^{12}$C$(\alpha,\gamma)$ reaction (as defined for example in \cite{Fow84}) can be deduced with high accuracy of approximately 12\%, using a global R-matrix analysis of existing data. We demonstrate that existing data do not permit the claimed 12\% accuracy, or even close to it, not withstanding sophisticated R-matrix analyses. We also raise strong doubts on the values of the astrophysical cross section factors quoted by Schuermann {\em et al.} \cite{Sch12}.

Two comments are in order from the outset. First, the ``rigorous data selection criteria" of Schuermann {\em et al.} \cite{Sch12} include ten data points shown with the measured 100\% (or more) error bars. Such data points should be considered as upper limits and as such they cannot be included in a rigorous chi-square analysis, since the contribution of an upper limit to chi-square cannot be rigorously evaluated. Most bothersome are the five such data points shown at low energies below 2 MeV (2 data points for $S_{E1}$ and 3 for $S_{E2}$). More significantly as we discuss below, other data included by Schuermann {\em et al.} \cite{Sch12} were shown in a later analysis (by this author \cite{Gai13}) to have similar large error bars (close to 100\%) which also render these data not useful for a rigorous chi-sqaure analysis.

Second, the data points below 2 MeV are essential for having high sensitivity to the contributions of the bound $1^-$ and $2^+$ states at 7.1169 and 6.9171MeV in $^{16}$O, respectively, that govern the value of the astrophysical cross section factors at stellar energies: $S_{E1}(300)$ and $S_{E2}(300)$, respectively. But the contribution to $\chi^2$ of these low energy data points is overwhelmed by the large number of high energy data points included by Schuermann {\em et al.}, leading to a reduced sensitivity to the low energy data. In addition the inclusion of higher energy data points raises additional question(s): for example concerning the sign of the interference of the $2^+$ at $E_{cm} = 2.68$ MeV that was shown to be positive and leading to an extrapolated $S_{E2}(300) = 62^{+9} _{-6}$ keVb \cite{Say12,Bru13}. This S-factor is 16\% lower than quoted by Schuermann {\em et al.} with a difference that is considerably larger than the quoted uncertainty of $S_{E2}(300)$. We note that the exclusion of high energy data points and consideration of data points only below 1.7 MeV was shown by this author \cite{Gai13} to lead to very different conclusions as we discuss below.

More troubling is the inclusion and in fact their reliance on the published data of Assuncao {\em et al.} \cite{Ass06}. We note from the outset that Brune {\em et al.} \cite{Say13} concluded several systematical problem with the data of Assuncao {\em et al.} \cite{Ass06}. These data \cite{Ass06} were re-analyzed by this author \cite{Gai13} and I revealed very large error bars of the extracted s-factors \cite{Gai13}, considerably larger than stated by Assuncao {\em et al.} \cite{Ass06}. 

We recapitulate here a few observation made in my chi-square analysis \cite{Gai13} of the data of Assuncao {\em et al.} \cite{Ass06}. For example even though no discernible peaks can be established in the gamma-ray spectra measured at $90^\circ -130^\circ$ shown in Fig. 6 of Assuncao {\em et al.} \cite{Ass06}, an angular distribution is claimed to have been measured at these backward angles at the indicated energy of $E_{\alpha,lab} = 1.850$ MeV. I showed that the angular distribution labeled as $E_{\alpha,lab}=1.900$ MeV (E = 1.340 MeV) \cite{Ass06} can be fitted with $S_{E2} \over S_{E1}$ values that vary by a factor of six with similar reduced $\chi^2$ values; see  Fig. 1 of Ref. \cite{Gai13}. My chi-square analysis of all data points measured by Assuncao {\em et al.} \cite{Ass06} below 1.7 MeV lead to $S_{E2} \over S_{E1}$ values that vary by a large factor without significant variation in chi-square. The resultanting uncertainties are shown in Fig. 2 of my paper \cite{Gai13} and they are close to 100\%. As  discussed above such data points must be considered as upper limits and the data of Assuncao {\em et al.} \cite{Ass06} cannot be included in a rigorous chi-square analysis. We emphasize that excluding the results of Assuncao {\em et al.} is not a matter of choice for ``data selection", rather it is dictate by standard considerations of a rigorous chi-square analysis. 

In the same paper I also demonstrated that the disagreement of the E1-E2 measured phase angle $(\phi_{12})$ with (theoretical) predictions, shown in Fig. 11 of Assuncao {\em et al.} \cite{Ass06}, is a violation of unitarity \cite{Gai13} and not just a mere disagreement with the prediction of R-matrix theory as suggested by them \cite{Ass06}. A violation of unitarity is a clear indication of serious problems with the data or the data analysis.

We conclude that the ``rigorous data selection" employed by Schuermann {\em et al.} \cite{Sch12} should not include the data of Assuncao {\em et al.} \cite{Ass06} and we remove the data of Assuncao {\em et al.} from the data sample analyzed by Schuermann {\em et al.} But since they already removed the data of Redder {\em et al.} \cite{Redder} and Ouellet {\em et al.} \cite{Ouellet} from their choice for the current precise data, the adjusted choice of data of $S_{E2}$ measured below 2 MeV includes only the data of Kunz {\em et al.} \cite{Kunz}. This is clearly an unacceptable situation that must be alleviated by new data measured at low energies and it cannot be remedied by sophisticated global R-matrix analyses. In the same time it is important to comment here that my analysis of the data of of Kunz {\em et al.} \cite{Kunz} agrees with Kunz {\em et al.} as stated in my paper \cite{Gai13}.

\begin{figure}
\begin{center}
 \includegraphics[width=5in]{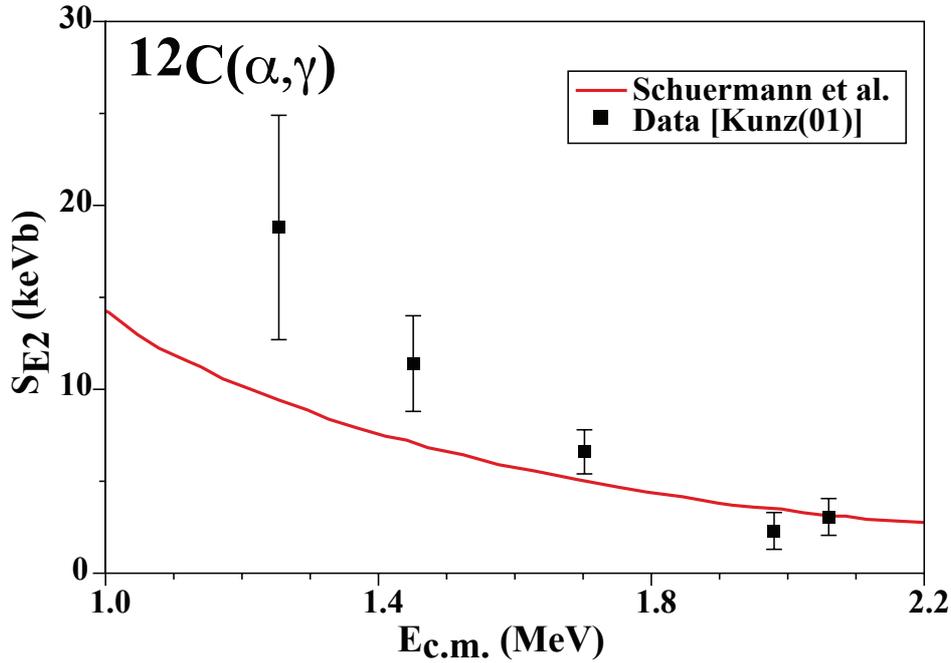}
 \end{center}
 \caption{\label{fig} (Color online) Comparison of the low energy portion of the R-matrix fit of Schuermann {\em et al.} \cite{Sch12} with the $S_{E2}$ data measured by Kunz {\em et al.} \cite{Kunz}. As discussed in the text the data of Assuncao {\em et al.} \cite{Ass06} have been removed from their ``rigorously selected data" leading to a major discrepancy with the remainder of the data (i.e. the Kunz {\em et al.} data) and specifically with the wrong predicted slope.}
\vspace{-0.5cm}
\end{figure}

Nevertheless the comparison of the low energy R-matrix curve shown  in Fig. 1, taken from their Fig. 4 \cite{Sch12}, reveals a disagreement by up to a factor of 2 with the adjusted choice of data sample used by Schuermann {\em et al.} (which does not include the data of Assuncao {\em et al.} and only includes the data of Kunz {\em et al.}, excluding the data points measured with 100\% error bars). More troubling is the fact that in comparison to the data their R-matrix fit has the wrong slope at low energies as shown in Fig. 1. Clearly these data points below 2 MeV are crucial for delineating the contribution of the bound $2^+$ state of $^{16}O$ and the slope is very important for an accurate extrapolation of $S_{E2}(300)$. 

We conclude that when considering the adjusted choice of the data sample used by Schuermann {\em et al.}, the value they quote for $S_{E2}$ cannot be substantiated and certainly we cannot support their claimed small uncertainty of the extracted $S_{E2}(300)$. We refer the reader to Ref. \cite{Gai13} for a complete chi-square analysis of all currently available data below 1.7 MeV that bifurcate and lead to ambiguity in the extracted value of $S_{E2}(300)$ =  $60 \pm 12$ or $154 \pm 31$ keVb.

Concerning the value of $S_{E1}(300)$: Schuermann {\em et al.} \cite{Sch12} consider the possibility of destructive interference of the 
$1^-$ sub-threshold state at 7.1169 MeV and the 2.42 MeV $1^-$ resonance in $^{16}$O and they claim ``the constructive solution is strongly favored and the destructive interference pattern has been rejected". This claim is based on the obtained $\chi^2_{cap}$ = 265 and 233 for $S_{E1}(300)$ = 7.9 and 83.4 keVb, respectively. When considering the large number of capture data points included by Schuermann {\em et al.}  (243) we conclude that the obtained $\chi^2$ difference of 32 is hardly significant to warrant the rejection of the destructive interference pattern [yielding the small $S_{E1}(300) = 7.9$ keVb]. 

Furthermore, since they state that a one sigma variation of the total s-factor leads to $\Delta \chi^2 \approx 21$ \cite{Sch12}, the obtained $\chi^2$ difference of 32 is not significant enough and certainly far from the usual five sigma used to substantiate a claim. Clearly both fits have reduced $\chi^2$ values that are close to unity, and the fit that leads to the small $S_{E1}(300) = 7.9$ keVb cannot be considered as ruled out by the data since the reduced $\chi^2$ is close to unity. As such the destructive interference cannot be ruled out with the certainty claimed by Schuermann {\em et al.} It is important to note that it is not sufficient to demonstrate the good fit for the large $S_{E1}(300)$ solution, but one must also rule out the small $S_{E1}(300)$ solution. This has not been achieved by Schuermann {\em et al.} and current data do not allow us to rule out the small $S_{E1}(300)$ solution, leading to the ambiguous value of $S_{E1}(300)$ \cite{Gai13}.

The rejection of the small $S_{E1}$ solution by Schuermann {\em et al.} based on $\chi^2$ consideration is made more doubtful since they included and relied on the data of Assuncao {\em et al.} which were found to have unrealistic small error bars \cite{Gai13}. This makes the minuscule difference in the reduced $\chi^2$ even more troublesome for rejecting the destructive interference pattern. 

Similarly the fit to the modern data that was published in a peer reviewed conference proceedings \cite{Ham05} (and was neglected by Schuermann {\em et al.}) states the numerical values: $S_{E1}(300)$ = 77.9 and 4.3 keVb with $\chi^2$ = 9.0 and 9.6, respectively, see Fig. 5 of \cite{Ham05}. Such a small difference in $\chi^2$ in of itself does not allow rejecting the small value solution and the need to re-evaluate the error bars quoted in Ref. \cite{Ass06,Ham05} weakens the possibility of rejecting the small s-factor solution using these modern data alone. We thus conclude that this modern gamma-ray data analysis \cite{Ham05} just the same as the analysis of Schuermann {\em et al.} \cite{Sch12} and previous data analyses \cite{Hale,Gia01} do not support ruling out the small $S_{E1} \approx 10$ keVb solution.

We conclude that a realistic evaluation of current data does not permit the determination of the astrophysical cross section factors with the 12\% accuracy suggested by Schuermann {\em et al.}, nor can we support their claimed values of $S_{E1}(300)$ and $S_{E2}(300)$. A more suitable conclusion is that both extrapolated $S_{E1}(300)$ and $S_{E2}(300)$ cross section factors are ambiguous with the values listed in my previous publication \cite{Gai13}. 

Unlike the strong claim of 12\% accuracy suggested by Schuermann {\em et al.} the observed large ambiguities justify and promote a new and different research effort to determine the astrophysical cross section factors with the required uncertainty of 10\% or better. Indeed proposals for determining the cross section at very low energy have been developed for the HI$\gamma$S gamma-ray facility in the USA  \cite{Gai}, and the newly constructed ELI-NP facility in Bucharest as shown in \cite{Gai14}. Measurements with gamma-beams are favored by the detailed balance factors of 50 - 100 and are made possible due to the anticipated high intensity ($10^9 \gamma /s$) that will allow a measurement at E$_{c.m.}$ = 1.0 MeV within two weeks of beam time. The design goal sensitivity of these measurements which is shown in Figs. 2 and 3 of \cite{Gai14}, promises to resolve the observed ambiguities in $S_{E1}(300)$ and $S_{E2}(300)$.

In closing we note that for example as shown in Fig. 1 of \cite{Wea93} the suggested value of the total astrophysical cross section factor quoted in \cite{Sch12} of $161 \pm 19  +8 -2$ keVb (i.e. a multiplicative factor of 1.61 as defined in \cite{Wea93}) is exactly at the boundary (170 keVb) where a 25 solar masses star is predicted to be oxygen rich (C/O $<$ 1) and thus skip the carbon burning stage and collapse to a black hole. Thus a resolution of the ambiguities in $S_{E1}(300)$  (approximately 10 or 80 keVb) and $S_{E2}(300)$ (approximately 60 or 154 keVb) noted in Ref \cite{Gai13} is essential for progress in stellar evolution theory.


\section{Acknowledgement}

I acknowledges helpful discussions with Mohammad W. Ahmed, Carl R. Brune and Hans O.U. Fynbo. Friendly communication from Lucio Gialanella and Frank Strieder led to a refined definition of the disagreement. This material is based upon work supported by the U.S. Department of Energy, Office of Science, Office of Nuclear Physics, under Award Number DE-FG02-94ER40870.





\bibliographystyle{elsarticle-num}
\bibliography{<your-bib-database>}







\end{document}